# "Evaluación de la Contaminación del Aire por Material Particulado PM2.5 en la ciudad del Cusco Respecto de los Índices de Calidad del Aire entre 2017 y 2018"

# " Assessment of Atmospheric Pollution by Particulate Matter PM2.5 in the city of Cusco respect to the Air Quality Indices between the years 2017 and 2018"

Bruce Stephen Warthon Olarte[2], Ivan Cesar Miranda Hankgo[2], Iván Ruben Quispe Ccolque[2], Rafael Eduardo Ponce Amanca[2], Victor Fernando Ramos Salcedo[2], Ariatna Isabel Zamalloa Ponce de Leon[2], Ruben Alfredo Tupayachi Latorre[2]. Julio Lucas Warthon Ascarza[1]

[1]*Departamento Académico de Física*
[2]*Escuela Profesional de Física*
[1,2.] *Facultad de Ciencias. Universidad Nacional de San Antonio Abad del Cusco (UNSAAC).*
*ORCID: 0000-0002-0985-1579*
*Correo electrónico: 160531@unsaac.edu.pe*

**Resumen**

En este artículo científico se evaluó la data de la contaminación del aire por material particulado $PM_{2.5}$ en diferentes lugares de la ciudad del Cusco respecto de los Índices de Calidad Ambiental (INCA) del Ministerio del Ambiente del Gobierno Peruano. Los resultados muestran que la contaminación del aire en la ciudad del Cusco es un problema ambiental de riesgo. Mas del 84% de los sitios monitoreados tiene calificación mala (101-500), el color correspondiente es anaranjado, este resultado muestra que el aire que respiran los ciudadanos del Cusco es de mala calidad y la población podría experimentar problemas de salud, la recomendación es evitar realizar ejercicios y actividades al aire libre. En el distrito de San Jerónimo Cusco la concentración promedio ha sido de 125 ug/m3 su equivalente en el intervalo INCA es >125, se encuentra dentro del color rojo correspondiente al umbral de cuidado, los efectos en la salud se describen como enfermedades pulmonares crónicas y cardiovasculares, la autoridad de salud debe declarar niveles de estado de alerta. Se ha concluido que según los INCA el aire de la ciudad del Cusco es mala y el inicio del valor de umbral del estado de cuidado (VUEC).

*Palabras clave: Contaminación del aire, material particulado $PM_{2.5}$, Índice de la Calidad Ambiental.*

**Abstract**

In this scientific article, the data on air pollution by PM2.5 particulate matter was evaluated in different locations in the city of Cusco with respect to the Environmental Quality Indexes (INCA) of the Ministry of the Environment of the Peruvian Government. The results show that air pollution in the city of Cusco is an environmental risk problem. More than 84% of the monitored sites have a bad rating (101-500), the corresponding color is orange. This result shows that the air that citizens of Cusco breathe is of poor quality and the population could experience health problems. The recommendation is to avoid outdoor exercises and activities. In the district of San Jerónimo Cusco, the average concentration has been 125 ug/m3, which corresponds to the INCA interval of >125, within the red color threshold for care. The health effects are described as chronic lung and cardiovascular diseases, and the health authority should declare levels of alert. It has been concluded that according to INCA, the air in the city of Cusco is of poor quality and falls within the threshold of the Care State Value (VUEC).

**Introducción**

La contaminación del aire es uno de los principales problemas ambientales a nivel mundial, las partículas en suspensión también conocidas como material particulado (PM), son los contaminantes más comunes que están presentes en la atmósfera. Estas partículas tienen un diámetro inferior a 2.5 micras (PM2.5) y son capaces de penetrar en los pulmones, causando problemas de salud como enfermedades respiratorias y cardiovasculares (World Health Organization, 2021). En los últimos años, la concentración de PM2.5 en la atmósfera terrestre ha sido objeto de estudio en diversas regiones del mundo con la finalidad de dar a conocer la problemática que existe debido al material particulado (PM). La Ciudad del Cusco viene experimentando un deterioro en la calidad del aire, los valores de las concentraciones de los contaminantes atmosféricos PM2.5 han sobrepasado los Estándares de Calidad del Aire en el distrito de San Jerónimo Cusco (OEFA, 2018), estos niveles conllevan riesgos en la salud pública. Según estudios epidemiológicos, la exposición prolongada a niveles elevados de PM2.5 está asociada con un aumento de enfermedades respiratorias crónicas, enfermedades cardiovasculares, cáncer de pulmón y otros problemas de salud (Pope et al., 2020). La contaminación del aire por PM2.5 se debe a varios agentes principalmente de origen antrópico como las emisiones de vehículos, la quema de combustibles fósiles, la actividad industrial, (Dockery & Pope, 1944).

La exposición a altos niveles de PM2.5 causó más de cuatro millones de muertes prematuras en 2016, representando cerca del 8% del total de las muertes en todo el mundo (Cohen et al., 2017). Desde 1970, se han realizado múltiples estudios para evaluar los niveles de PM2.5 en distintas áreas geográficas y para comprender sus efectos sobre la salud humana (Brunekreef & Holgate, 2002; Pope & Dockery, 2006).

Los valores altos de concentración de PM2.5 en lugares urbanos se relacionan con la quema de combustibles fósiles, la actividad industrial, las emisiones de polvo de la construcción y la quema de biomasa para calefacción y cocina (Zhang, 2017). También es importante destacar, según Jiang (2017), que las condiciones meteorológicas pueden influir en la concentración de PM2.5 en el aire. Por ejemplo, la estabilidad atmosférica y la falta de viento pueden llevar a una acumulación de contaminantes en una zona específica, lo que puede aumentar la concentración de PM2.5 en el aire.

La evaluación de la concentración de $PM_{2.5}$ es importante en la gestión de la calidad del aire y la identificación de áreas críticas que requieren medidas de control y mitigación, y se lleva a cabo para evaluar los niveles de contaminación del aire y los posibles riesgos para la salud, se realizó una investigación que involucró el monitoreo de $PM_{2.5}$ en diferentes lugares de la ciudad del Cusco mediante un muestreador de material particulado de alto volumen HIVOL de marca Ecotech con certificación de la agencia de protección ambiental US-EPA.

La ciudad del Cusco presenta particularidades geográficas y climáticas que pueden influir en la concentración de PM2.5. Además, la topografía local y la vegetación pueden afectar la dispersión de PM2.5 en el aire (Zhang et al., 2010). Su ubicación en la Cordillera de los Andes, a una altitud de más de 3,300 metros sobre el nivel del mar, y su clima seco y soleado, pueden contribuir en la concentración de contaminantes atmosféricos (Instituto Nacional de Estadística e Informática, 2018). Debido a estas particularidades geográficas se ha realizado la evaluación de la concentración del PM2.5 usando los datos medidos a través de una red de monitoreo automatizada conocida como Sistema de monitoreo Shelter del Centro de Investigación de Energía y Atmósfera - UNSAAC en la ciudad del Cusco el cual cuenta con cinco analizadores automáticos de gases de efecto invernadero y un equipo muestreador de PM, durante el período de 2017 a 2018.

La ciudad del Cusco se halla en un valle interandino de la zona sur oriental denominado valle del Huatanay, en la sierra del Perú ubicada entre los 3,244 y 3,700 m.s.n.m, a 13° 30' 45" latitud Sur y a 71° 58' 33" longitud Oeste. En esta región, García M.. (2019) destaca la presencia de otros factores meteorológicos que pueden contribuir a la concentración de PM2.5 en la ciudad de Cusco. Debido al legado histórico, la ciudad de Cusco recibe cientos de miles de turistas anualmente (Reporte Regional de Turismo - Cusco, 2021), el sector que genera mayor emisión de contaminantes atmosféricos es parque automotor (PRAL, 2006). El material particulado PM2.5. fue medido por DIGESA (2013), OEFA (2018), los resultados muestran que este parámetro ha sobrepasado los ECA establecidos por el MIANAM (DS 003-2017-MINAM). El segundo sector que contamina más la cuenca atmosférica (PRAL, 2006) es el sector de la fabricación de ladrillos artesanales.

Para evaluar la contaminación atmosférica en la ciudad del Cusco se ha considerado los Índices de Calidad Ambiental para Aire (INCA); el rango numérico está comprendido entre 0 y 100, es adimensional y está relacionado a los Estándares de Calidad Ambiental de Aire (MINAN, 2016). Este índice se obtiene a partir de la medición de diversos contaminantes, tales como PM2.5, PM10, NO2, O3 y SO2, y se presenta en una escala de PM2.5 que varía de 0 a 500 (MINAN, 2016). Los valores del índice de calidad del aire más altos implican una mayor presencia de contaminantes en el aire, lo que aumenta los riesgos para la salud y el medio ambiente (US EPA, 2016). En ese sentido, evaluar la concentración de $PM_{2.5}$ en la ciudad del Cusco es importante por su relación directa con la salud pública, el medio ambiente, el patrimonio natural y cultural de la ciudad, el conocimiento sobre los niveles de contaminación y los INCA permitirían una planificación urbana y toma de decisiones asertiva.

Referente a la metodología se aplicó la evaluación espacial, medida del promedio de las concentraciones de material particulado, grafico mediante Power Bi, para la concentración de $PM_{2.5}$, los mapas satelitales proporcionan información valiosa para la toma de decisiones en la planificación del transporte, la gestión de la calidad del aire y la protección de la salud pública. Esta herramienta puede ayudar a identificar áreas críticas de contaminación y diseñar estrategias de control de la contaminación. Cohen et al. (2017) destacan la importancia de la evaluación de la contaminación del aire en la salud pública. Para llevar a cabo la evaluación espacial de la contaminación atmosférica, se emplean herramientas de georreferenciación y análisis espacial como los Sistemas de Información Geográfica (SIG), los cuales facilitan la visualización y análisis de datos geoespaciales. Estos sistemas permiten identificar las zonas de mayor riesgo de contaminación, las cuales pueden ser mapeadas y analizadas para detectar patrones espaciales y tendencias en la distribución de la contaminación en diferentes áreas geográficas. (Jerrett et al., 2010). Según Yang et al. (2022) sugiere evaluar la distribución espacio temporal de las concentraciones de $PM_{2.5}$ durante un periodo de tiempo.

**Material y Métodos**

**Figura 1**

*Shelter para Monitoreo Ambiental (UNSAAC)*

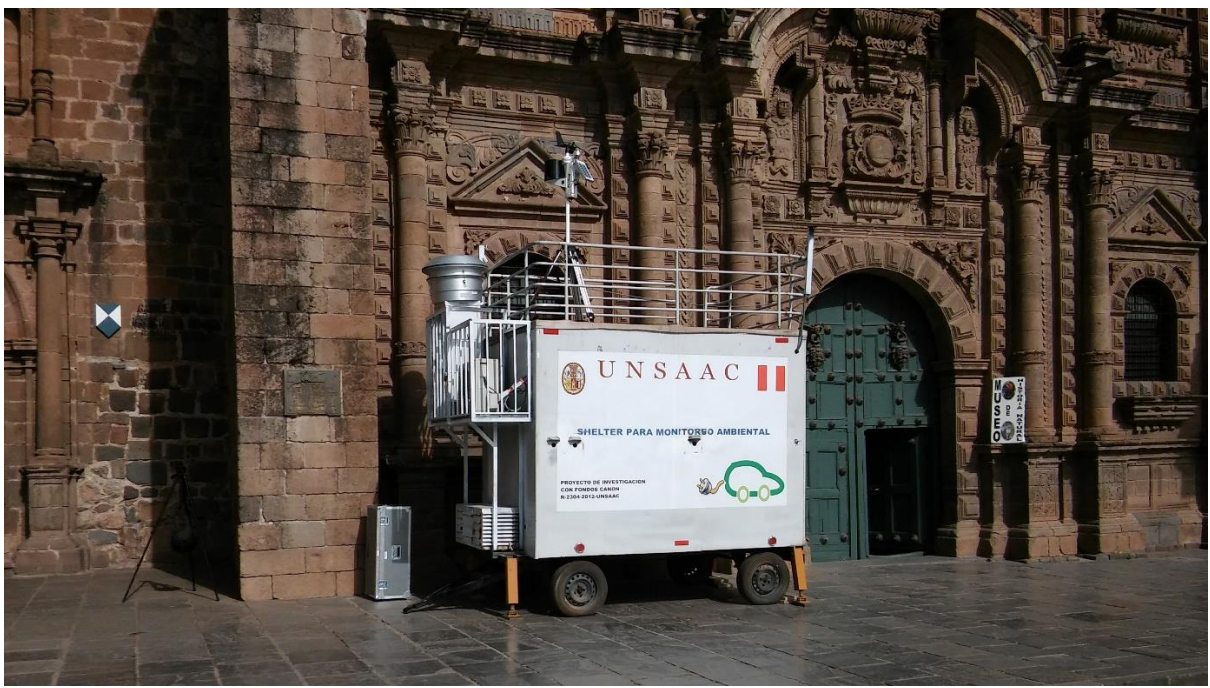

*Fuente: Elaboración propia, 2017.*

*Equipo Muestreador:*

**Figura 2**

*Muestreador de material particulado de alto volumen (Hi-Vol 3000)*

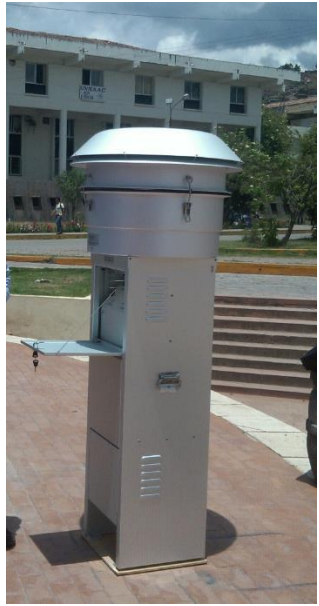

*Fuente: Libro Energía limpia y cero emisión, 2017.*

*Material de Muestreo:*

Para recolectar el material particulado pululante en el medio ambiente es preciso usar filtros de recolección, por ello en esta investigación se usan los filtros de cuarzo debido a que la muestra recolectada es representativa y recomendada a nivel mundial por su porosidad.

**Figura 3**

*Filtro de cuarzo sobre Hi-Vol en la ciudad de Cusco*

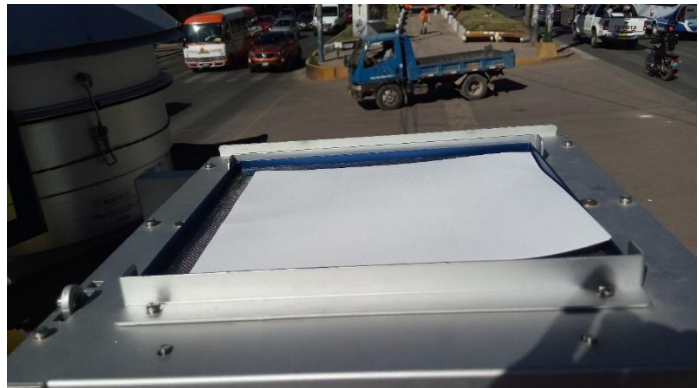

*Fuente: Elaboración propia, 2017.*

*Método de muestreo:*

En función a los filtros empleados, se ha aplicado la gravimétrica (método analítico cuantitativo), esto debido a la variación de masa que existe entre un valor inicial y un valor final para estudiar la concentración de material particulado.

*Procedimiento para recolección de muestra:*

El muestreo se hizo de acuerdo al protocolo establecido por el MINAM:

1. Los sitios de muestreo no fueron aleatorios, la selección de los lugares se definió considerando los siguientes criterios: cercanía a vías de tránsito vehicular y fábricas de ladrillo. En ambos casos el tránsito de peatones se realiza simultáneamente a las emisiones de contaminantes atmosféricos.
2. El shelter de monitoreo ambiental se transportó a los diferentes lugares de medición, el muestreador de material particulado (Hi-Vol) se ubicó en la parte posterior del shelter.
3. Un día previo a la medición se pesó cuatro veces la masa de un filtro de cuarzo obteniendo su masa inicial promedio (laboratorio de energía y atmósfera, UNSAAC), seguidamente se colocó dentro de una campana de desecación durante 24 horas. Al día siguiente se extrae el filtro para medir su masa cuatro veces y la obtención del promedio que se considera como valor referencial inicial (mi), seguidamente se transporta el filtro en una caja hermética hacia la ubicación donde se encuentra el Hi-Vol.
4. Se coloca el filtro dentro del Hi-Vol y se enciende el equipo durante 24 horas continuas
5. Al finalizar del periodo de muestreo se retira el filtro del Hi-Vol para transportarlo al laboratorio, se midió la masa final cuatro veces obteniendo su promedio, seguidamente se coloca dentro del desecador de vidrio durante 24 horas, luego de este lapso se midió cuatro veces la masa final obteniendo el promedio final (mf).
6. Mediante el método gravimétrico se determinó la masa del material particulado
7. La concentración del material particulado obtenido en el filtro se determino mediante las fórmulas establecidas en el protocolo establecido por el MINAM y EPA.

**Figura 4**

*Retiro de filtro de Cuarzo luego de su exposición al ambiente durante 24 horas consecutivas.*

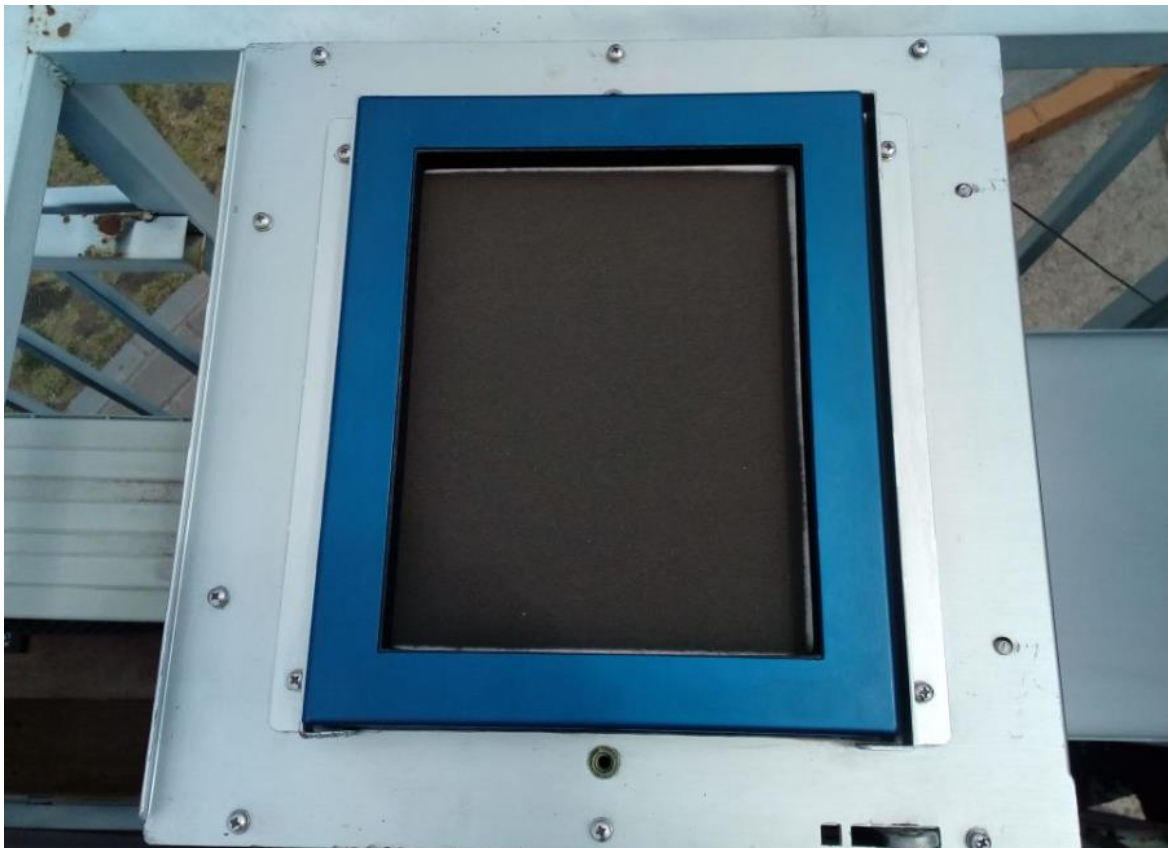

*Fuente: Elaboración propia, 2017.*

**Figura 5**

*Codificación de filtros de cuarzo correspondientes al año 2017.*

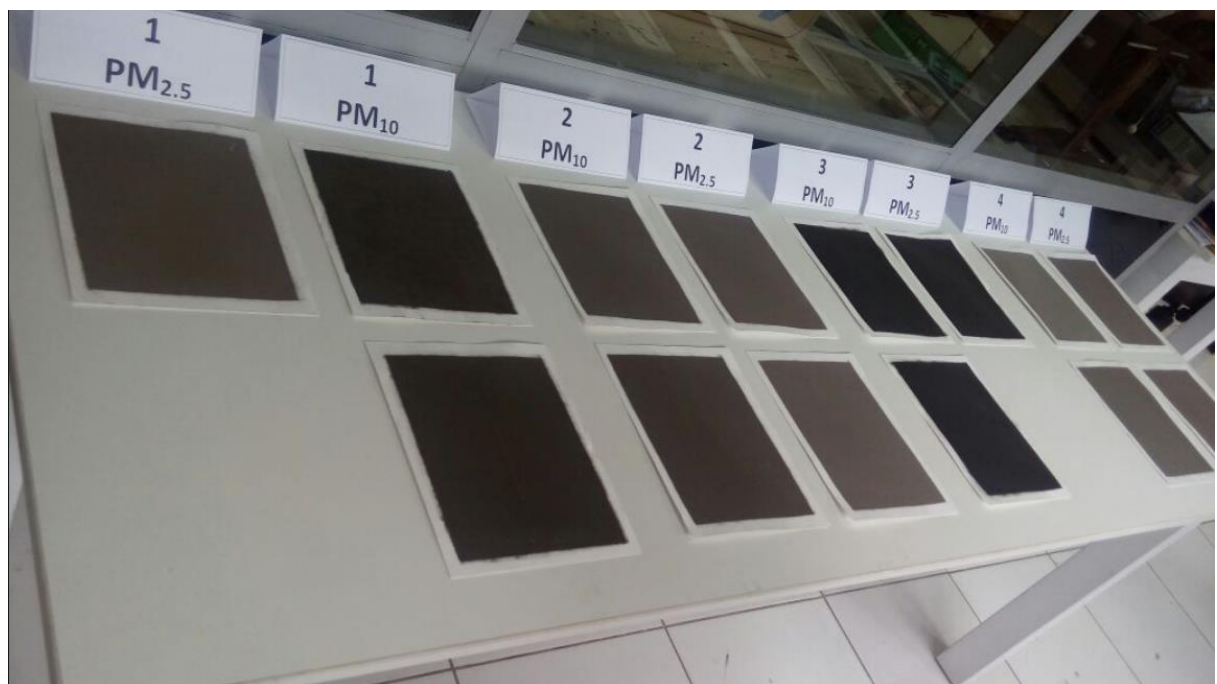

*Fuente: Elaboración propia, 2017.*

**Resultados**

**Tabla 1**

*Datos recogidos entre los años 2017 y 2018, periodo de medición 24 horas mediante el equipo Hi-Vol.*

| Punto de monitoreo | Distrito | Lugar de monitoreo | Promedio de PM $_{2.5}$ | Mes de monitoreo | Año de monitoreo |
|---|---|---|---|---|---|
| 1 | Cusco | Campus Universitario: Facultad de Educación | 65 | Mayo | 2017 |
| 2 | San Sebastián | APV Paraíso de Fátima | 153.2 | Junio / Agosto | 2017 |
| 3 | Cusco | Campus Universitario, puerta número 05 (multired) | 70.27 | Setiembre | 2017 |
| 4 | Cusco | Plaza de Armas | 81.87 | Octubre | 2017 |
| 5 | Cusco | Colegio Clorinda Matto de Turner | 79.23 | Diciembre | 2017 |
| 6 | Wanchaq | Local del Municipio Distrital | 71.68 | Diciembre | 2017 |
| 7 | Wanchaq | Centro de Salud (CLAS) contiguo a la estación de bomberos | 118.41 | Enero | 2018 |
| 8 | Cusco | Plaza San Francisco (cerca a pileta) | 52.81 | Enero / Febrero | 2018 |
| 9 | Cusco | San Blas | 20.19 | Marzo | 2018 |
| 10 | Cusco | Plazoleta Limacpampa | 68.48 | Marzo | 2018 |
| 11 | Cusco | Plazoleta Pumacchupan | 117.47 | Abril | 2018 |
| 12 | Santiago | Centro de salud, Belen Pampa (CLAS) | 98.67 | Mayo | 2018 |
| 13 | San Jerónimo | Centro de Salud aledaño al local de la policia (CLAS) | 105.66 | Junio / Agosto | 2018 |

*Fuente: Elaboración propia, 2023.*

Según la tabla 1, la evaluación de promedios y valores máximos de concentración de

PM$_{2.5}$ en gráficos de barras se muestra en la Figura 6.

**Figura 6**

Valores de concentración promedios de PM$_{2.5}$ en cada punto de medición en la ciudad del Cusco.

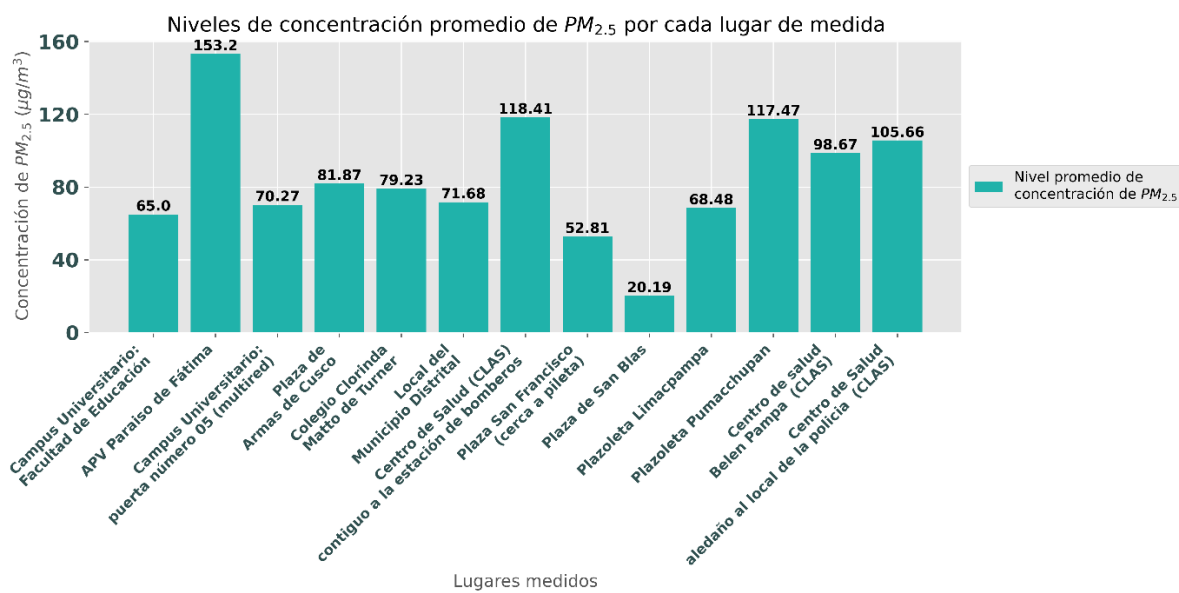

*Fuente: Elaboración propia en Python, 2023.*

**Tabla 2**

Índice de Calidad Ambiental referente a Material Particulado PM$_{2.5}$

| | Material particulado (PM$_{2.5}$) promedio 24 horas | | |
|---|---|---|---|
| Color | Calificación | Intervalo del INCA | Intervalo de concentraciones ($\frac{\mu g}{m^3}$) |
| Verde | Buena | 0 - 50 | 0 - 12.5 |
| Amarillo | Moderada | 51 - 100 | 12.6 - 25 |
| Naranja | Mala | 101 - 500 | 25.1 - 125 |
| Rojo | Umbral de Cuidado | >500 | >125 |

*Fuente: RM-N°-181-2016-MINAM.*

El índice de calidad de aire (INCA) dado para cada punto de monitoreo que se presenta en la tabla 3, fue elaborado usando los parámetros establecidos en la Resolución Ministerial N° 181-2016-MINAM. Estos parámetros se muestran en la ecuación (1).

$$I(PM_{2.5}) = [PM_{2.5}] * \frac{100}{25} \cdots (1)$$

**Tabla 3**

Cálculo de Índice de Calidad de Aire para los puntos de monitoreo de material particulado PM$_{2.5}$ de la tabla 1.

| Punto de monitoreo | Concentración PM$_{2.5}$ ($\frac{\mu g}{m^3}$) | INCA | Color | Calificación |
|---|---|---|---|---|
| 1 | 65.00 | 260.00 | Naranja | Mala |
| 2 | 153.20 | 612.80 | Rojo | Umbral de cuidado |
| 3 | 70.27 | 281.06 | Naranja | Mala |
| 4 | 81.87 | 327.46 | Naranja | Mala |
| 5 | 79.23 | 316.91 | Naranja | Mala |
| 6 | 71.68 | 286.71 | Naranja | Mala |
| 7 | 118.41 | 473.63 | Naranja | Mala |
| 8 | 52.81 | 211.23 | Naranja | Mala |
| 9 | 20.19 | 80.77 | Amarillo | Moderada |
| 10 | 68.48 | 273.93 | Naranja | Mala |
| 11 | 117.47 | 469.89 | Naranja | Mala |
| 12 | 98.67 | 394.67 | Naranja | Mala |
| 13 | 105.66 | 422.62 | Naranja | Mala |

*Fuente: Elaboración propia, 2023.*

**Figura 7**

Mapa de distribución de la concentración promedio de material particulado PM$_{2.5}$ por cada punto de monitoreo según datos de la tabla 1.

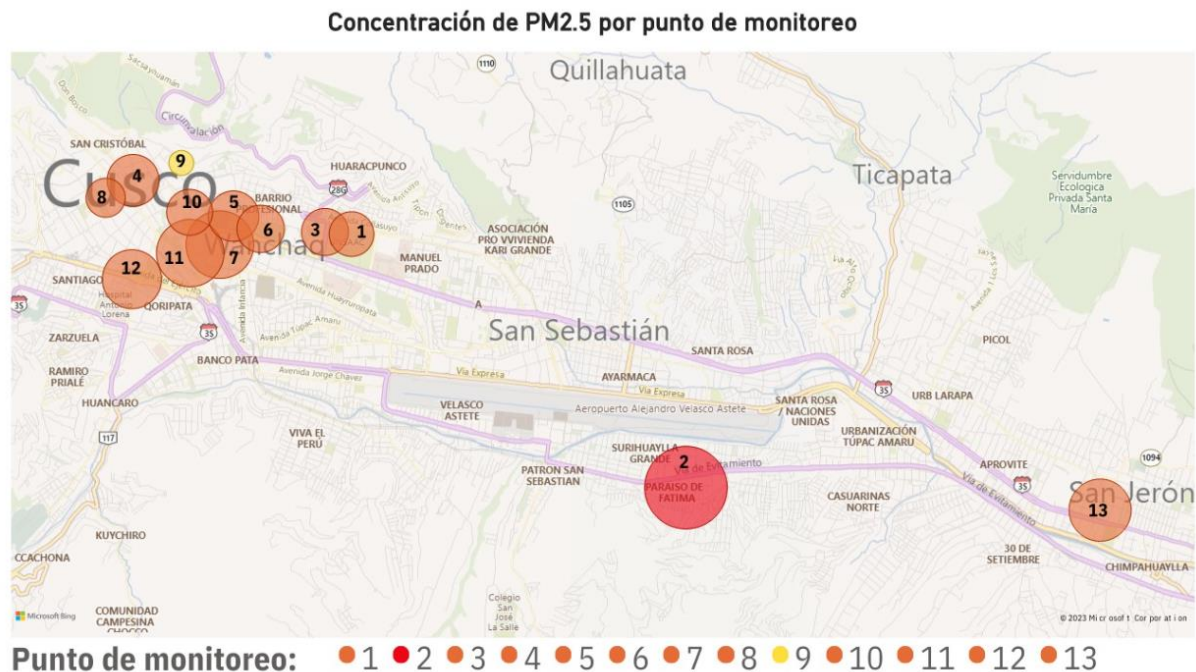

*Fuente: Elaboración propia en Power Bi, 2023*

**Discusión**

El análisis de los datos de concentración promedio de PM$_{2.5}$ en la ciudad del Cusco durante el período de estudio (2017-2018) revela patrones de estudio interesantes en la distribución geográfica de este contaminante atmosférico. Utilizando el programa Power Bi, se generó una representación visual (Figura 7) de la cantidad promedio de concentración de

$PM_{2.5}$ por zonas, mediante círculos cuyo tamaño representa la magnitud de la concentración en cada punto de monitoreo, lo cual permite una evaluación visual de la contaminación del aire en la ciudad.

Se observa que el punto 2, ubicado en el APV Virgen de Fátima, presenta los niveles más elevados de concentración de $PM_{2.5}$ en comparación con los demás puntos de monitoreo. Estos altos niveles de concentración sugieren que la contaminación por $PM_{2.5}$ en esta zona podría estar relacionada con las ladrilleras cercanas, ya que la afluencia de vehículos en esta área es muy baja. Por otro lado, el punto 9, ubicado en San Blas, muestra los niveles más bajos de concentración de $PM_{2.5}$ en comparación con los demás puntos de monitoreo. Esto podría deberse a la difícil geografía de la zona, lo cual limita la fluidez del tráfico vehicular en esta área.

En los demás puntos de monitoreo, se observa que los niveles de concentración de $PM_{2.5}$ son esperables debido a su cercanía a la Av. La Cultura, que es la avenida más transitada en la ciudad del Cusco, o a otras calles y avenidas céntricas con mayor flujo de tráfico vehicular, a diferencia del punto 9. Esto sugiere que la actividad vehicular es una de las principales fuentes de contaminación por $PM_{2.5}$ en la ciudad del Cusco.

Estos resultados concuerdan con estudios previos que han identificado el tráfico vehicular y las emisiones de las ladrilleras como importantes fuentes de contaminación del aire por $PM_{2.5}$ en la ciudad del Cusco. Sin embargo, es importante destacar que la concentración de $PM_{2.5}$ también puede verse afectada por otros factores, como las condiciones meteorológicas, la topografía del terreno y la estacionalidad, entre otros. Por lo tanto, es necesario realizar un análisis más detallado e integrado de estos factores para comprender plenamente los patrones de contaminación del aire en la ciudad del Cusco.

Además, es relevante destacar que los niveles de concentración de $PM_{2.5}$ en la ciudad del Cusco durante el período de estudio se encuentran por encima de los límites establecidos por las normas de calidad del aire, lo cual indica una preocupante situación de contaminación atmosférica en la ciudad. La exposición a niveles elevados de $PM_{2.5}$ se ha asociado con una serie de efectos adversos para la salud humana, incluyendo enfermedades respiratorias, cardiovasculares y otras condiciones crónicas. Por lo tanto, es necesario tomar medidas adecuadas para reducir la contaminación del aire por $PM_{2.5}$ en la ciudad del Cusco, con el fin de proteger la salud de la población.

En este contexto, es fundamental implementar estrategias de mitigación, como la promoción del uso de transporte público y medios de transporte no motorizados, así como la mejora de la tecnología de las ladrilleras para reducir las emisiones de contaminantes. De igual manera, es necesario concientizar a la población sobre la importancia de reducir su huella de carbono y adoptar prácticas sostenibles en su vida cotidiana.

Adicionalmente, es relevante establecer políticas y regulaciones más estrictas en cuanto a la calidad del aire y promover la participación activa de la comunidad en la toma de decisiones relacionadas con la gestión de la contaminación atmosférica. También es importante fomentar la investigación y monitoreo continuo de la calidad del aire para tener una comprensión más precisa de las fuentes y niveles de contaminación en la ciudad del Cusco.

En resumen, el análisis de los datos de concentración promedio de $PM_{2.5}$ en la ciudad del Cusco revela patrones de contaminación atmosférica relacionados principalmente con el tráfico vehicular y las emisiones de las ladrilleras. Estos resultados resaltan la necesidad de implementar medidas de mitigación y concientización para reducir la contaminación del aire y proteger la salud de la población. Es fundamental adoptar enfoques integrados y multidisciplinarios para abordar este problema y garantizar un ambiente saludable y sostenible en la ciudad del Cusco.

**Conclusiones**

El análisis de los datos de concentración promedio de $PM_{2.5}$ en la ciudad del Cusco durante los años 2017 y 2018 demuestran que la contaminación atmosférica en cada sitio de medición está relacionada principalmente con el tráfico vehicular y las emisiones de las ladrilleras ubicadas en la ciudad del Cusco. La concentración de $PM_{2.5}$ más alta en el año 2017 se midió la APV Virgen de Fátima, este punto de monitoreo se encuentra cerca al sector de ladrilleras. El punto de monitoreo ubicado en San Blas muestra los niveles más bajos de concentración de $PM_{2.5}$, posiblemente debido al bajo tránsito vehicular.

Mas del 84% de los puntos de monitoreo demuestran valores de concentración de $PM_{2.5}$ elevados, según el INCA su calificación es mala y en el caso de la APV Virgen de Fátima su calificación es de Umbral de Cuidado, en el caso de San Blas su calificación es moderada; por consiguiente, el predominio de la calificación INCA es mala en la ciudad del Cusco; lo cual sugiere implementar estrategias de mitigación de manera urgente.

Es necesario concientizar a la población sobre la importancia de reducir su huella de carbono y adoptar prácticas sostenibles en su vida cotidiana. Además, se requiere establecer políticas y regulaciones más estrictas en cuanto a la calidad del aire, así como promover la participación activa de la comunidad en la toma de decisiones relacionadas con la gestión de la contaminación atmosférica. La investigación y monitoreo continuo de la calidad del aire también son esenciales para obtener una comprensión más precisa de las fuentes y niveles de contaminación en la ciudad del Cusco.

**Declaración de conflicto de intereses**

Los autores declaran no presentar ningún tipo de conflicto de intereses